\documentclass[12pt,a4paper]{article}
\usepackage[utf8]{inputenc}     
\usepackage[T1]{fontenc}        
\usepackage{lmodern}            
\usepackage[british]{babel}
\usepackage{amsmath}
\usepackage{amssymb}
\usepackage{amsfonts}
\usepackage{setspace}           
\usepackage{graphicx}
\title{%
What is Time? -- Thoughts of a Physicist\\
{\normalsize Contribution to the ``Symposium on Time''}\\
\vspace{-1.5ex}
{\normalsize of the European Psychoanalytical Federation,
Brussels, April 2022}%
}
\author{Gernot M\"unster%
\thanks{E-mail: munsteg@uni-muenster.de}\\
University of M\"unster,\\
Institute for Theoretical Physics\\
and Center for the Philosophy of Science,\\
D-48149 M\"unster, Germany}
\date{17th March 2022}
\newcommand{\be}{\begin{equation}}
\newcommand{\ee}{\end{equation}}
\newcommand{\bea}{\begin{eqnarray}}
\newcommand{\eea}{\end{eqnarray}}

\begin{document}
\maketitle
%
\section{Introduction}
The notion of time plays a fundamental role in human experience, thinking and feeling. 
Our image of the world is inconceivable without the concept of time. 
So, what is time?

With this question the sixth chapter of Thomas Mann’s ``The Magic Mountain'' begins: 
``What is time? A mystery – insubstantial and almighty. A prerequisite of the world of
appearances, a motion, coupled and mingled with the being of bodies in space
and their movement. But would there be no time, if there were no motion? 
No motion if no time? Just ask! Is time a function of space? Or the other way around?
Or are both identical? Go ahead, ask!''\,\cite{Mann}

For a physicist, time is a concept of prime importance. So, let us ask about the nature of time.
This, however, is more a question for philosophers. Not being a philosopher, allow me to
skate on thin ice for a moment. In the literature you find quite different answers.
According to Wittgenstein\,\cite{Wittgenstein}, ``What is time?'' is not a meaningful question, 
but a misleading one. ``Time'' is not the name of an object. On the other hand, Kant gives an
answer in his transcendental philosophy: according to him, time is a pure form of sensible 
intuition (\emph{reine Form der Anschauung}).
``Time is a necessary representation that grounds all intuitions. In regard to appearances
in general one cannot remove time, though one can very well take the appearances away from
time. Time is therefore given \emph{a priori}. In it alone is all actuality of appearances 
possible.''\,\cite{Kant}

Natural scientists often express themselves in a more casual way. The physicist John A.\ Wheeler,
the Ph.\,D.\ adviser of Richard Feynman,
liked to characterise time in the way he found it in a graffito in the men's room of the 
Pecan Street Cafe in Austin, Texas, in 1976: 
``Time is nature's way to keep everything from happening all at once.''\,\cite{Wheeler}

The phenomenon of time leads to a number of interesting questions:
Does time have a beginning, or an end? Is ``time travel'' possible? Can time cyclically close in
a circle? How does the difference between past and future come about?
Which of these questions can be answered by physics?

A distinction concerning the concept of time, which is often discussed in the philosophical
literature about time, has been made by the Cambridge philosopher 
J.~M.~E.\ McTaggart (1866--1925)\,\cite{McTaggart}. 
He introduced the notions of A-series and B-series for the temporal ordering of events.
In the A-series, events are divided into those belonging to the past, to the present, 
and to the future. The classification continuously changes: events are first in the future, then
in the present, and finally remain in the past. Time is of course ordered in the A-series:
for each pair of unequal times, $t_1$ and $t_2$, either $t_1$ is earlier than $t_2$ or vice versa.
The A-series allows tensed propositions, like ``we had a reception yesterday evening''.
In the B-series, temporal positions are just linearly ordered with the earlier/later relation.
It does not make use of the notions of past, present and future. Only tenseless propositions,
like ``we had a reception in the evening of April 8th'',
are possible in the B-series. The most significant difference between the two series
is the fact that in the A-series there is a ``now'', whereas this concept is missing in the B-series.
In the philosophy of time, those who consider the A-series or the B-series to be more fundamental 
are called A-theorists or B-theorists, respectively.
Philosophers discuss whether future and past are real, or whether they exist, and they
debate about the status of the present (see e.\,g.\ \cite{Poidevin}).
We shall, however, not delve into these metaphysical questions here.
\begin{figure}[hbt]
\vspace{.8cm}
\centering
\includegraphics[width=0.7\textwidth]{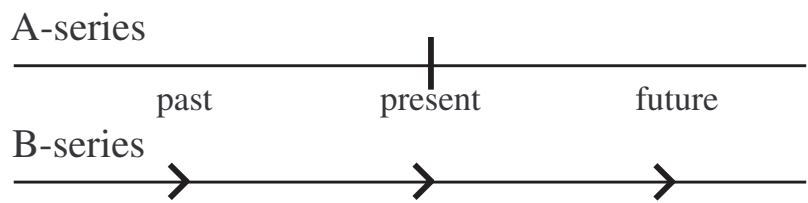}
\parbox[t]{0.8\textwidth}{%
\caption{McTaggart's A-series and B-series.}%
\label{fig:AB-series}%
}
\end{figure}

Time, the experience of time, and the measurement of time have philosophical, biological,
physiological, psychological, sociological and physical aspects, which are viewed differently 
in the different disciplines. For example, in physiology the subjectively experienced ``moment'' 
(the critical time interval $T_c$) covers a period of about 30 msec, 
and the ``subjective presence'' (the mental presence time $T_p$) has a duration of several
seconds\,\cite{Gruesser-Poeppel},
while in physics the presence is understood as the sharp boundary between past and future. 

In the following we shall deal with a few aspects of time in the context of physics.

\section{Time in physics}
Einstein is said to have answered the question asked at the beginning of this paper with: 
``Time is what the clock shows.''
What at first glance seems like a flippant answer is in fact an expression of his years of 
struggle over the fundamentals of space and time. The operational approach to time, reflected in the 
quote, was the key to setting up the theory of relativity. 
The associated success and progress in understanding time is based on a 
retreat, a simplification, in that within the framework of physics the subjective elements 
are disregarded and we content ourselves with the measurable.
This notion of time, measured by clocks, is called ``physical time'' or ``objective time''.

On the other hand, there is ``phenomenological time'' or ``subjective time'', 
which is the time that we see passing in our individual experience.
Depending on our particular state of being, phenomenological time might appear to run
faster or slower than physical time.
Also, our experience of events and processes includes the modal structure of time,
corresponding to the A-series, with its past, present and future.

In contrast to that, physics does not know the ``now''.
Why is this so?
In general, physics is not interested in single events or processes. It is not interested
in the fall of Newton's apple or of my teacup. Instead, it is interested in regularities, which
it tries to describe in terms of laws, e.\,g.\ Galileo's law of falling bodies.
Such a law is supposed to be valid (within its scope of application and within a certain 
precision) universally and at all times. 
Therefore, references to singular points in time, and in particular to the ``now'', are
not admitted in physical laws.
It should be mentioned that in exceptional cases reference to particular dates is made
in physics, like in astrophysics, where, for example,
the detection of gravitational waves on 14th September 2015 and
Tycho Brahe's supernova of 11th November 1572
are considered. Nevertheless, these dates are referred to in a tenseless way.
So, in general, physical statements about temporal instances are made according to the
B-series.

Are physical time and phenomenological time in contradiction to each other?
Are the A-series and B-series incompatible?
I do not think so.
Every account of temporal processes or events has its origin in human perception.
Expressing experiences in terms of phenomenological time or physical time,
using the A-series or the B-series
means to employ different representations of the same reality.
Objective time is obtained by an abstraction of subjective time, based on
an intersubjective agreement about the meaning of clocks.

\section{Measurement of time}
Time, more precisely time intervals, are measured with clocks. What is a clock? 
Clocks are characterised by periodic processes that are used to define periods of time.
In the course of history, time measurement developed from the rotation of the earth to water clocks 
and wheel clocks to more modern technical constructs. 
A milestone was the chronograph by John Harrison from 1759.
The development continued to today's quartz clocks and atomic clocks. 
The precision of the currently available atomic clocks has given rise to the valid 
definition of the fundamental unit of time by means of atomic processes: 
One second corresponds to 9,192,632,770 periods of the radiation emitted by
the transition between the two hyperfine levels of the ground state of atoms of caesium 133.

Is not a vicious circle lurking here? 
Is not the definition of time units via periodic processes, whose uniformity is assumed, circular? 
Would not the old definitions of the hour and the second via the rotation of the earth 
be just as viable as the one using modern atomic clocks?
In this spirit, Percy W.~Bridgman (1882--1961), winner of the 1946 Nobel Prize in physics, 
proposed to define physical concepts and units in an operational way\,\cite{Bridgman}.
The meaning of a scientific term would then be given by a specific measurement method.
Consequently, there would be different seconds: an ``earth second'', a ``quartz clock second'',
a ``caesium clock second'' etc.
When I was young, I read in a newspaper about the -- at that time -- most accurate atomic clock.
Its precision was also specified, and I wondered how it could have been determined, if no
more accurate clock existed for comparison.  
Indeed, in the operational framework it would not make sense to ask for the precision of an 
atomic clock, if time units are defined by means of this clock.

The actual meaning of time and definition of time units in physics is, however, different.
The concept of time is theory-laden. 
It does not have its meaning by virtue of any particular periodic process, 
which is defined to be uniform, 
but within an entire system of description of nature, which should be consistent in itself and simple. 
The laws of physics contain time as a parameter, often denoted $t$.
For example, in Newtonian mechanics the motion of a free particle is described by the equation
of motion $\ddot x = 0$ and its solution
$x(t) = v t$, where $v$ is the velocity. The time parameter $t$ is involved in the laws
of electrodynamics, relativity, quantum theory, etc.
It is a theoretical term that finds its meaning within the framework of a theoretical description 
of a domain of phenomena.
It is specified in such a way that the laws of physics take a simple form.
Imagine, for example, that by contrast $t$ would be defined in terms of the rotation of the earth.
This would be logically possible, but
in this case the motion of a free particle would not be uniform, and the various laws of
physics would take rather complicated and impractical forms, 
since the rotation of the earth is not uniform with respect to these other phenomena.
Instead, time is an idealised concept, which is tied to the phenomena and experiments by means of
models, idealisations and bridge principles.
Time intervals and units are then best represented by clocks that currently run most uniform 
according to their physical description.

\section{Time's arrow}
The term ``arrow of time'' means the fact that time has an inherent direction 
that distinguishes the past from the future. It becomes apparent in our experience 
in everyday life and in the sciences. 
The directionality of time has been put into words very nicely by Schiller\,\cite{Schiller}:
\begin{verse}
Threefold is the march of time:\\
While the future slow advances,\\
Like a dart the present glances,\\
Silent stands the past sublime.

\hspace{10ex}
{\footnotesize Friedrich Schiller, Proverbs of Confucius}
\end{verse}
C.\,F.~von Weizsäcker\,\cite{Weizsaecker} has summed up this fact in the succinct statement
``the past is factual, the future is possible.''
This structure of time appears to be inextricably linked to the concept of time.
Why is that?
Kant again had a deep insight into this.
In our lives we constantly have experiences.
Gaining experience, however, means learning from the past for the future.
So this structure of time, the difference between past and future, is a necessary condition 
for the possibility of experience.

Physics is an empirical science; it is based on experience.
The arrow of time is thus a prerequisite for physics.
Therefore, at first glance it appears reasonable to assume 
that physics cannot justify the arrow of time.
Is that really the case? I do not think so.
The circle that arises when physics searches for a physical reason for the arrow of time 
does not have to be vicious.
It is a matter of semantic consistency\,\cite{Weizsaecker} whether the physical justification for the 
arrow of time, expressed in the language of physics, 
and the arrow of time as a prerequisite for physics fit together.
So let us turn to the issue of an eventual physical justification for time's arrow.
There is a lot of literature about this, but I will only mention the books \cite{Halliwell,Zeh,Davies}.

The fundamental laws of physics that describe particles and their interactions are 
time-reversible.\footnote{There is an exception to this in the microphysical laws for 
the weak interactions of elementary particles (CP violation), which, however, does not matter 
for the present discussion.}
Each process may also run backwards in time.
But the processes taking place in nature are obviously not time-reversible:
a cup that falls to the ground shatters into many pieces; the reverse process has never been observed. 
How does that come about?
Where does this arrow of time come from, 
which does not originate from the fundamental laws of nature?

On closer inspection, one can distinguish several arrows of time:
\begin{enumerate}
\item
psychological: our memory is directed towards the past and not towards the future,
\item
thermodynamic: according to the 2nd law of thermodynamics, entropy always increases
(I leave out the details and necessary prerequisites for this law here),
\item
electrodynamic: radiation propagates outwards from the source as time progresses, and does
not flow in concentrically,
\item
quantum theoretical: the changes of state in the measurement process of a quantum system
are irreversible,
\item
cosmological: the universe is expanding.
\end{enumerate}

Detailed inspection shows that the time's arrows 1 to 4 are connected to each other.
The irreversible processes associated with them take place on the basis of probability laws,
which describe a progression from ordered to more disordered states.
Thereby, the overall disorder of the systems under consideration increases.
Counterintuitive as it might appear, this also applies to our memories.
To remember some event means that it has left traces
in our memory, which are realised by certain neuronal structures.
Traces in general represent remnants of irreversible processes, which are accompanied by
an increase of the total disorder (entropy).

In physics, entropy is a measure of the disorder of a system.
The time arrow can then be traced back to the fact that an ordered initial state 
(with low entropy) has been present.
While the laws are time-reversible, the initial conditions lead to irreversibility of processes. 
The question of time's arrow therefore leads to the question of the origin of the ordered 
initial state. 
There is still no generally accepted answer to this, but there is a plausible hypothesis 
that the origin is based in cosmology, and the expansion of the universe acts as a kind 
of ``master arrow of time''.

A remark is in order here. 
On earth we often observe the emergence of ordered structures in apparent contradiction to 
the increase in disorder (entropy). 
This contradiction is illusory: in open systems, complexity can grow locally 
at the expense of disorder in the outside world\,\cite{Prigogine}. 
Evolution does not conflict with thermodynamics.

\section{Theory of relativity}
It is not possible for a physicist to talk about time without taking the theory of relativity
into account. It was established in one of the five famous works from Einstein's 
``annus mirabilis'', 1905, entitled \glqq Zur Elektrodynamik bewegter 
Körper\grqq \ (``On the Electrodynamics of Moving Bodies'').
The title sounds inconspicuous and rather technical. But this work is about the very nature 
of space and time. In it, Einstein overthrows concepts of space and time that had been
thousands of years old. 

Difficulties in electrodynamics and optics led Einstein to investigate the fundamentals 
of the concepts of space and time. From 1902 to 1909 he worked in the patent office in Bern. 
He wrote to his friend Conrad Habicht\,\cite{Habicht}: ``After eight working hours there 
are still eight hours for miscellaneous, and a whole Sunday.''
The result of this was the theory of relativity, which brought with it a radical new way 
of thinking about space and time.
Consequences of the theory are:
\begin{enumerate}
\item
Relativity of simultaneity,
\item
Relativity of time,
\item
Relativity of lengths.
\end{enumerate}

Let me make a few remarks about the relativity of time. What is meant with that?
The passage of time, more precisely the running of clocks, depends on how the observer 
and the clock move relative to each other. A clock moving at a certain speed relative to us 
runs slower than a clock at rest. 
So does running keep you young? Well, the effect is extremely small for ordinary speeds.
The factor by which the moving clock of a cyclist slows down is 1.00000000000000017, 
which is equivalent to 1 second in 200 million years.
Only at speeds comparable to the speed of light of 299,792.458 km/sec does the effect 
become significant. At 90\% of the speed of light the factor is 2.3 and at 99\% of the 
speed of light it is already 7.1.

It is important that not only artificial clocks are affected, but all processes, including 
biological ones. It is time itself, the passing of which is relative. The relativity of time 
is often illustrated by the so-called twin paradox: a twin leaves earth in a spaceship, 
which travels into space at high speed and returns several years later. While the twin who 
remained on earth has aged to an old man, his much less aged sister climbs out of the spaceship. 
Although the realisation of this story is far beyond today's possibilities, the effect has been 
confirmed experimentally with the help of atomic clocks in airplanes.

The effects resulting from the theory of relativity are by no means esoteric fantasies of 
scientists, but play a role in many areas of today's physics and technology. 
An example is the GPS (Global Positioning System). Without taking into account the theory 
of relativity, an error of 10 km would add up in the GPS devices every day,
and some of you would have missed today's venue by a long way!

\section{Simultaneity}
The relativity of simultaneity is a consequence of the theory of relativity, which is of
crucial significance for our understanding of time. Surprisingly, it has received rather
little attention among philosophers. Exceptions are, for example, H.~Putnam\,\cite{Putnam} and
works cited in \cite{Wuethrich}.

\begin{figure}[hbt]
\centering
\includegraphics[width=0.7\textwidth]{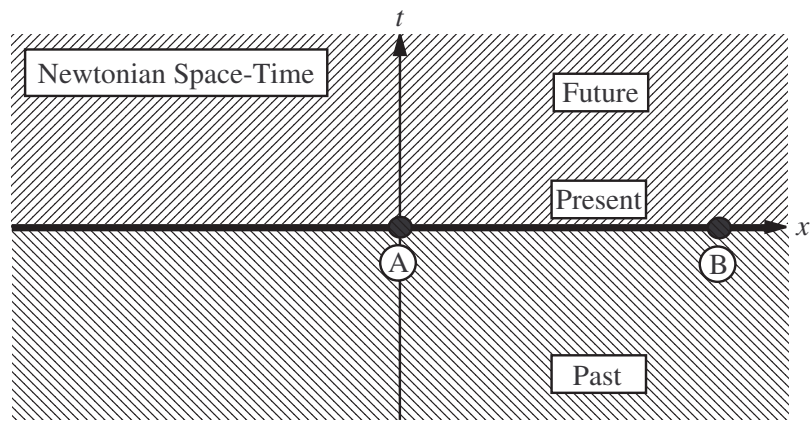}
\parbox[t]{0.8\textwidth}{%
\caption{Newtonian space-time. Shown are the time axis and one spatial axis.
The global present is indicated by a thick horizontal line.}%
\label{fig:Newton}%
}
\end{figure}

In physics before Einstein, the idea of an absolute time prevailed, 
which is equally universally valid and measurable for every observer. 
In his ``Philosophiae Naturalis Principia Mathematica'' from 1687 Newton writes: 
``Absolute, true, and mathematical time, of itself, and from its own nature, flows equably 
without relation to anything external.''
Absolute time implies that a concept of global simultaneity is possible.
If in Newtonian space-time an event A takes place at a certain location $x_A$ and time $t_A$,
then an event B, happening at any other location at the same time $t_A$, is simultaneous
with A. The existence of an absolute notion of simultaneity allows to define the concepts of
future, present and past unambiguously, see Fig.~\ref{fig:Newton}.
What is now for me here, is now for everybody elsewhere.

\begin{figure}[hbt]
\centering
\includegraphics[width=0.7\textwidth]{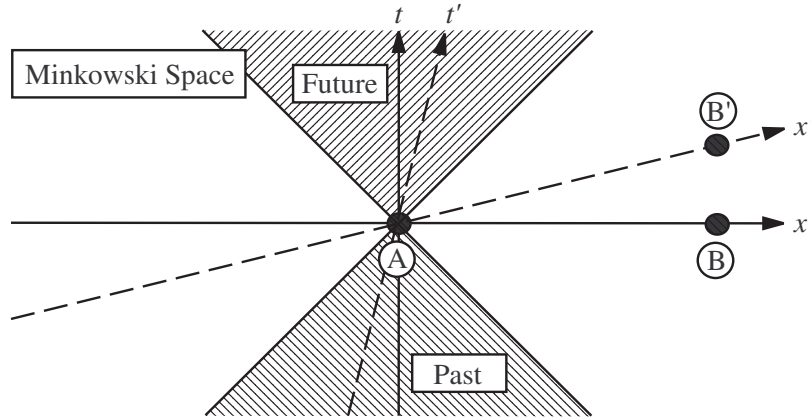}
\parbox[t]{0.8\textwidth}{%
\caption{Minkowski space. Shown are the time axis and one spatial axis.
Note that for better visibility of relativistic effects, units are chosen such 
that propagation of light takes place on diagonal lines.
So, if 1 cm on the $t$-axis corresponds to 1 sec, then 1 cm on the $x$-axis corresponds to
300.000 km.
Causal past and causal future are abbreviated as past and future.
Depending on the velocity of an observer at A, events B or B$'$ are Einstein-simultaneous
with A.}%
\label{fig:Minkowski}%
}
\end{figure}

The relativistic situation is substantially different from the Newtonian one.
The geometry of space-time is represented by the so-called Minkowski space,
see Fig.~\ref{fig:Minkowski}.

The global past and future of an event A in Newtonian space-time are now replaced by two regions
called ``causal past'' and ``causal future'', depending on A.
(Choosing $x_A=0$ and $t_A=0$ for simplicity, they are specified by the conditions 
$t \leq - |x|/c$ and $t \geq + |x|/c$, respectively, where $c$ is the speed of light.)
The Newtonian present of A has become a whole region of space-time between past and future,
the points in it being called space-like relative to A.

Which points in Minkowski space are simultaneous with A?
Consider the following situation in Fig.~\ref{fig:Minkowski}. 
Physicist Bob works in a distant laboratory. At a certain time,
he is looking for the result of a quantum physical experiment. This event is denoted B.
Let us assume that the result of the experiment is not predetermined.
For Bob, its outcome is open before B, and factual afterwards.
At point A philosopher Alice wonders how the experiment of physicist Bob went.
For her, the question is whether event B has already taken place or not.
That amounts to the question of simultaneity of A and B.

Looking at Fig.~\ref{fig:Minkowski}, it seems as if A and B are simultaneous.
But Einstein's revolutionary insight was that there is no unambiguous concept of simultaneity.
As he showed, one can indeed define a kind of simultaneity, called ``Einstein-simultaneity'',
which, however, depends on the velocity of the observer A and is thus a relative notion.
I refer to Einstein's beautiful 
book \textit{Relativity: The Special and the General Theory}\,\cite{Einstein},
where he explains these issues in a generally understandable way
by means of moving trains.
If the $x$- and $t$-axes in Fig.~\ref{fig:Minkowski} correspond to the rest frame of Alice,
then the points on the $x$-axis, including B, are Einstein-simultaneous to A.
As mentioned above, this notion depends on the motion of Alice.
If she walks through her office, her rest frame is represented by the $x'$- and $t'$-axes,
obtained from the previous coordinates by a Lorentz transformation.
(For better visibility, the angles are exaggerated in the figure.)
The event in Bob's laboratory, which is Einstein-simultaneous to A, is now indicated
by B$'$. In everyday situations the difference will be extremely small.
But for the sake of principle, consider Bob's laboratory to be located in the
Andromeda galaxy. If Alice moves with 3 km/s towards him, the event B$'$ is later than
B by a considerable amount of 9 days.

\begin{figure}[hbt]
\centering
\includegraphics[width=0.7\textwidth]{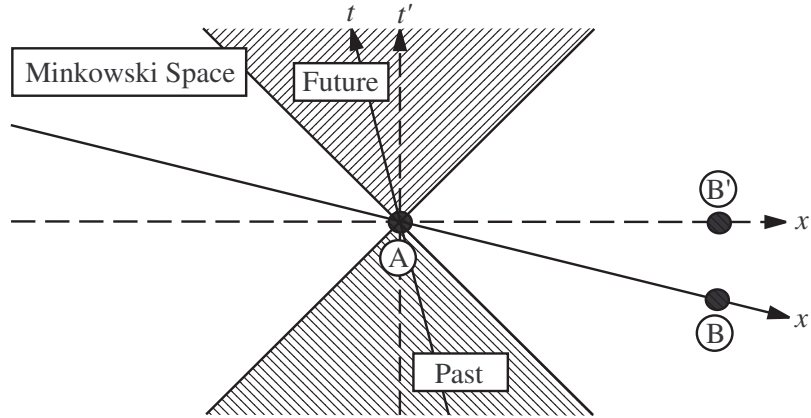}
\parbox[t]{0.8\textwidth}{%
\caption{Minkowski space. The same situation as in Fig.~\ref{fig:Minkowski}
is drawn differently, but equivalently.}%
\label{fig:Minkowski2}%
}
\end{figure}

A remark is in order. Fig.~\ref{fig:Minkowski} might suggest that the $x$-$t$-frame is 
privileged because of the 90$^\circ$ angle between its axes.
But this is an optical illusion. This orthogonality does not have a meaning in Minkowski
geometry. The same situation can be displayed in an equivalent manner
as in Fig.~\ref{fig:Minkowski2}.

We have to see that for events at Bob's place, being simultaneous with event A does not have an 
inherent meaning.
So we have to realise that there is no unambiguous way for Alice to consider one of Bob's times as
``now'', simultaneous with her ``now''.

The inevitable conclusion of this consideration is that for Alice, now thinking about 
Bob's experiment, there is no way to give a meaning to statements about its status of
being completed or being still open.
This is a severe challenge for our understanding of reality, and for A-theorists in particular.
I consider this to be the most serious problem in the philosophy of time.

Einstein appears to have been aware of the consequences of the relativity of simultaneity.
He probably arrived at a deterministic worldview, in which there is no room for an open future.
Shortly before his death in 1955, in a letter to the family of his deceased friend
Michele Besso\,\cite{Besso}, he wrote ``For us believing physicists, the distinction 
between past, present and future only has the meaning of an illusion, though a stubborn one.''

%

\begin{thebibliography}{99}
%
\bibitem{Mann}
\glqq Was ist die Zeit? Ein Geheimnis, - wesenlos und allmächtig. 
Eine Bedingung der Erscheinungswelt, eine Bewegung, verkoppelt und vermengt dem Dasein der Körper im 
Raum und ihrer Bewegung. Wäre aber keine Zeit, wenn keine Bewegung wäre? 
Keine Bewegung, wenn keine Zeit? Frage nur! Ist die Zeit eine Funktion des Raumes? 
Oder umgekehrt? Oder sind beide identisch? Nur zu gefragt!\grqq\\
Thomas Mann, \textit{Der Zauberberg}, S.~Fischer Verlag, Frankfurt am Main, 2002;\\
English translation by me.
%
\bibitem{Wittgenstein}
L.~Wittgenstein, \textit{Das blaue Buch}, 1970.
%
\bibitem{Kant}
\glqq Die Zeit ist eine notwendige Vorstellung, die allen Anschauungen zum Grunde liegt. 
Man kann in Ansehung der Er\-schei\-nun\-gen überhaupt die Zeit selbst nicht aufheben, ob man zwar 
ganz wohl die Er\-schei\-nun\-gen aus der Zeit wegnehmen kann. Die Zeit ist also a priori gegeben. 
In ihr allein ist alle Wirklichkeit der Erscheinungen möglich.\grqq\\
I.~Kant, \textit{Kritik der reinen Vernunft, Der transzendentalen Ästhetik Zweiter Abschnitt, 
Von der Zeit}, 1781;
English translation: \textit{Critique of Pure Reason}, P.~Guyer and A.~W.~Wood (tr.),
Cambridge University Press, 1998.
%
\bibitem{Wheeler}
J.~A.~Wheeler, \textit{Time Today}, in \cite{Halliwell}.
%
\bibitem{Halliwell}
J.~J.~Halliwell, J.~P\'{e}rez-Mercader, and W.~H.~Zurek (eds.),
\textit{Physical Origins of Time Asymmetry}, Cambridge University Press, 1994.
%
\bibitem{McTaggart}
J.~M.~E.~McTaggart, \textit{The unreality of time}, Mind 17 (1908) 457--474.
%
\bibitem{Poidevin}
\textit{Questions of Time and Tense}, R.~Le Poidevin (ed.), Clarendon Press, Oxford, 1998.
%
\bibitem{Gruesser-Poeppel}
O.-J.~Grüsser, \textit{Zeit und Gehirn}; E.~Pöppel, \textit{Erlebte Zeit und die Zeit überhaupt};
both in: \textit{Die Zeit -- Dauer und Augenblick}, H.~Gumin and H.~Meier (eds.), Piper Verlag,
München, 1989.
%
\bibitem{Bridgman}
P.~W.~Bridgman, \textit{The Logic of Modern Physics}, Macmillan Co., New York, 1927.
%
\bibitem{Schiller}
\glqq
Dreifach ist der Schritt der Zeit:\\
Zögernd kommt die Zukunft hergezogen,\\ 
Pfeilschnell ist das Jetzt entflogen,\\ 
Ewig still steht die Vergangenheit.\grqq\\
Friedrich Schiller, Sprüche des Konfuzius;\\
English translation: \textit{The Poems of Schiller, Complete: Including All His Early Suppressed Pieces,
Attempted in English by Edgar Alfred Bowring},
John W. Parker \& Son, London, 1851.
%
\bibitem{Weizsaecker}
C.~F.~von Weizsäcker, \textit{Die Einheit der Natur}, Carl Hanser Verlag, München, 1971.
%
\bibitem{Zeh}
H.~D.~Zeh, \textit{The Physical Basis of the Direction of Time}, Springer, Berlin, 1992.
%
\bibitem{Davies}
P.~C.~W.~Davies, \textit{The Physics of Time Asymmetry}, University of California Press, 1974.
%
\bibitem{Prigogine}
I.~Prigogine, \textit{Time, Structure and Fluctuations}, Nobel Lecture, 8 December 1977.
%
\bibitem{Habicht}
\glqq Bedenken Sie, daß es im Tag neben den acht Stunden Arbeit noch acht Stunden Allotria 
und noch einen Sonntag gibt.\grqq\\
Letter from Albert Einstein to Conrad Habicht, 1905, 
in: \textit{The Collected Papers of Albert Einstein,
Volume 5: The Swiss Years: Correspondence, 1902--1914},
M.~J.~Klein, A.~J.~Kox, and R.~Schulmann (eds.),
Princeton University Press, 1994.
%
\bibitem{Putnam}
H.~Putnam, \textit{Time and Physical Geometry}, The Journal of Philosophy 64 (1964) 240--247.
%
\bibitem{Wuethrich}
C.~Wüthrich, \textit{The fate of presentism in modern physics},
in: \textit{New Papers on the Present -- Focus on Presentism},
R.~Ciuni, K.~Miller, and G.~Torrengo (eds.),
Philosophia Verlag, Munich, 2013, p.~91--131.
%
\bibitem{Einstein}
A.~Einstein, \textit{Relativity: The Special and the General Theory},
Dover Publications, New York, 2001;
original: \textit{Über die spezielle und die allgemeine Relativitätstheorie},
Vieweg, Braunschweig, 1917.
%
\bibitem{Besso}
\glqq Für uns gläubige Physiker hat die Scheidung zwischen Vergangenheit, Gegenwart und Zukunft 
nur die Bedeutung einer, wenn auch hartnäckigen, Illusion.\grqq\\
Letter of Einstein to Vero and Beatrice Besso, 21 March 1955,
in: \textit{Albert Einstein -- Michele Besso. Correspondance 1903--1955},
P.~Speziali (ed.), Hermann, Paris, 1972, p.~537--538.
%
\end{thebibliography}
\end{document}